\documentclass[sigconf,nonacm]{acmart}

\usepackage{amsmath}
\usepackage{booktabs}

\hypersetup{pdfpublisher={Authors}}

\begin{document}
\title{Modeling Preference Dependencies with Directed Acyclic Graphs for Narrative-Driven Recommendation}
\author{Mingzhe Lu}
\author{Jian Qin}
\author{Ruize Qin}
\author{Jinhao Liu}
\author{Yunpeng Li}
\author{Hang Su}
\renewcommand{\shortauthors}{Lu et al.}

\makeatletter
\renewcommand{\@mkauthors}{%
  \global\setbox\mktitle@bx=\vbox{%
    \unvbox\mktitle@bx
    \medskip
    \centering
    {\large\bfseries
      Mingzhe Lu\textsuperscript{1,2}\quad
      Jian Qin\textsuperscript{1,2}\quad
      Ruize Qin\textsuperscript{3}\quad
      Jinhao Liu\textsuperscript{1,2}\quad
      Yunpeng Li\textsuperscript{1,2,*}\quad
      Hang Su\textsuperscript{4,*}\par}
    \medskip
    {\normalsize
      \textsuperscript{1}Institute of Information Engineering, Chinese Academy of Sciences, Beijing, China\par
      \textsuperscript{2}School of Cyber Security, University of Chinese Academy of Sciences, Beijing, China\par
      \textsuperscript{3}Faculty of Engineering and IT, The University of Melbourne, Melbourne, Australia\par
      \textsuperscript{4}Independent Researcher\par}
    \smallskip
    {\small\textsuperscript{*}Corresponding authors: \href{mailto:liyunpeng@iie.ac.cn}{\texttt{liyunpeng@iie.ac.cn}}; \href{mailto:hangsucuhk@gmail.com}{\texttt{hangsucuhk@gmail.com}}.\par}
    \bigskip}}
\makeatother

\begin{abstract}
Narrative-driven recommendation allows users to express multifaceted needs through free-form text where stated preferences naturally qualify and condition one another. Existing methods typically treat narratives as flat keyword combinations or unconstrained generative queries, failing to capture how complex preferences hierarchically refine and contextualize each other. We propose the Preference Dependency Graph Framework (PDG-F), a narrative-driven recommendation framework centered on explicit preference dependencies. Specifically, PDG-A (PDG-Agent) first parses unstructured requests to organize user preferences and their mutual qualifications into a directed acyclic Preference Dependency Graph (PDG). Through this graph, candidate evidence from the open Web is acquired via directed path retrieval, and final recommendations are determined by topological path ranking. Comprehensive experiments across three diverse domains spanning English and Chinese contexts confirm that PDG-F consistently achieves superior recommendation performance over competitive baselines, open-domain retrieval agents, and frontier language models. Further diagnostic analyses show that our dependency-guided retrieval markedly raises the attainable candidate-pool ceiling, demonstrating the fundamental value of modeling preference dependencies in narrative-driven recommendation.
\end{abstract}

\maketitle

\section{Introduction}

\textit{Relevance to Web.} Traditional recommender systems on Web platforms largely rely on historical interactions or structured profiles to approximate user intent~\cite{adomavicius2005toward,koren2009matrix}. However, these signals often fall short as modern Web users increasingly turn to free-form natural language narratives~\cite{jannach2021conversational,kang2017understanding,li2018deep}. While the emergence of large language models (LLMs) offers new opportunities to interpret such narratives~\cite{wu2024llmrec,geng2022p5,sanner2023large}, retrieving candidates from open Web resources and evaluating their suitability require preserving how users' preferences condition and constrain one another. Against this backdrop, we study narrative-driven recommendation (NDR) in open-web environments, aiming to provide personalized recommendations by deciphering rich, unstructured user requests~\cite{bogers2017defining,afzali2021pointrec}.

Existing efforts have explored different paradigms to transform narratives into recommendation evidence. Early work primarily relied on explicit clues. For example, Bogers and Koolen~\cite{bogers2018combining} combined example items with narrative context. Eberhard et al.~\cite{eberhard2019evaluating} used annotated genres and keywords for post-filtering. Subsequent work further explored automated element extraction to reduce manual annotation~\cite{eberhard2024computing}. Another line explores semantic matching between narratives and items through embedding representations~\cite{eberhard2020embedding}. Recent work further explores the potential of directly generating recommendations from narratives with LLMs~\cite{he2023zeroshot,eberhard2025llmndr}. To further empower LLMs with external evidence, Zhu et al.~\cite{zhu2025crag} introduce collaborative retrieval to provide behavioral signals. Shi et al.~\cite{shi2025ocg} apply open-domain retrieval and attribute completion to rank candidates. Kanchi et al.~\cite{kanchi2026bookforyou} utilize generated plot tags for fine-grained reranking. However, preferences expressed in narratives often mutually qualify each other~\cite{partee1984compositionality}, a dimension still insufficiently modeled in existing methods.

User preferences in narratives are not merely flat collections of independent elements, but rather hierarchical and dependency structures formed through conditional qualifications. Consider the following movie recommendation request:
\begin{quote}
\itshape
``I'd like fantasy with romance as a subplot, rather than the main attraction. [\ldots] I need believable individual characters and a relationship shown through conversation, humour, and handling disagreements, not mostly sex.''
\end{quote}
In this query, ``subplot'' constrains the prominence of romance within the overarching narrative, while conversation, humour, and handling disagreements further qualify the specific manner in which this relationship unfolds. Neglecting such qualifying dependencies can easily cause a film that superficially covers fantasy, romance, and humor to be falsely identified as a match, such as \emph{The Lord of the Rings}. Yet the humor belongs to camaraderie instead of romance, failing the qualification. In contrast, \emph{Howl's Moving Castle} develops the romance through bickering and everyday interaction, satisfying the requested dynamic. This contrast reveals the necessity of explicitly modeling qualification and dependency structures among narrative preferences to guide the retrieval and evaluation of Web evidence for recommendation.

We propose PDG-F (Preference Dependency Graph Framework), a narrative-driven recommendation framework comprising a graph construction agent, directed path retrieval, and topological path ranking. Specifically, the graph construction agent, PDG-A, extracts diverse elements from the narrative description and unfolds them into discrete nodes through a two-layer structure of perspective and instantiation. On this basis, PDG-A evaluates the qualification relationships between node pairs within the context of the entire narrative to form directed edges. After resolving cycles, it generates the concrete preference for each current node in topological order conditioned on the preference context of its ancestral nodes, allowing upstream preferences to be progressively refined with each additional qualification to complete the preference dependency graph (PDG). Relying on this PDG, we first acquire candidates and their associated evidence from the open Web via directed path retrieval, and then compute candidate scores through topological path ranking, ultimately delivering recommendation results that better fit the specific preferences of the user.

The main contributions of this work are summarized as follows.
\begin{itemize}
  \item We propose PDG-F (Preference Dependency Graph Framework), a narrative-driven recommendation framework that organizes user qualifying constraints into a progressively refined directed dependency structure, providing a principled structural representation for complex composite preferences.
  \item We develop a structure-guided decision mechanism built directly upon the constructed PDG (Preference Dependency Graph), comprising directed path retrieval to acquire candidate evidence from the Web under specific qualification constraints and topological path ranking to compute candidate scores, thereby unifying evidence gathering and final recommendation under a coherent structure.
  \item We conduct comprehensive experiments across three distinct domains encompassing movies, books, and anime in both English and Chinese contexts, demonstrating that PDG-F consistently achieves state-of-the-art performance against competitive baselines as well as modern commercial retrieval engines alongside frontier models and LLM agents, with ablations and diagnostic analyses verifying the efficacy of dependency modeling.
\end{itemize}

\section{Related Work}
Translating natural narratives into accurate item recommendations hinges on effectively understanding and operationalizing the multifaceted needs woven into unstructured text. Guided by varying depths of textual interpretation and representation paradigms, existing efforts have broadly evolved across three paradigms: explicit feature matching, representation learning, and LLM-based recommendation with evidence augmentation.

Explicit feature matching focuses on distilling concrete recommendation signals from unstructured narratives through signal integration, request element extraction, and narrative trope generation. Early work by Bogers and Koolen~\cite{bogers2018combining} combined narrative context with referenced example items to exploit the complementarity between explicit exemplars and descriptive intent. To obtain more specific recommendation cues, Eberhard et al.~\cite{eberhard2024computing} automated the extraction of request elements into discrete recommendation cues, while Kanchi et al.~\cite{kanchi2026bookforyou} generated narrative trope tags from book descriptions to incorporate specific story patterns into candidate reranking. However, these methods primarily organize recommendation around feature matching, still lacking explicit representations for the modifier and qualification relationships among preference elements.

Representation learning approaches compute relevance by learning semantic correspondences between narrative requests and items, expanding from embedding-based matching~\cite{reimers2019sentence,karpukhin2020dense} to task adaptation, data augmentation, and general-purpose representation transfer. Early work by Eberhard et al.~\cite{eberhard2020embedding} leveraged embedding representations to analyze the predictive utility of distinct request elements, while Chaganty et al.~\cite{chaganty2023investigating} adopted contrastive learning to train dual-encoder retrieval models between conversational context and songs. Addressing the sparsity of narrative supervision, Mysore et al.~\cite{mysore2023mint} utilized LLMs to synthesize narrative queries from user interactions, providing supervision for dense retrieval models. To better capture domain semantics, Epure et al.~\cite{epure2024learning} fine-tuned models to align natural language requests with high-level track descriptions. Broadening the scope to general recommendation and search, Attimonelli et al.~\cite{attimonelli2025zero} validated the zero-shot capabilities of general-purpose text embeddings. Although these efforts substantially advance semantic relevance modeling, they still lack explicit characterization of the qualification scope and inter-dependencies among diverse preference criteria.

LLM-based methods leverage powerful language understanding and reasoning capabilities to interpret narrative intents, evolving from direct generation relying on internal parametric knowledge toward evidence-augmented paradigms that integrate external retrieval and structured attributes~\cite{lewis2020retrieval}. Eberhard et al.~\cite{eberhard2025llmndr} systematically evaluated different models and prompting strategies, benchmarking the feasibility and boundaries of directly generating recommendations from parametric memory. To alleviate the limitations of static parametric knowledge, subsequent efforts focus on equipping LLMs with external grounding evidence: Zhu et al.~\cite{zhu2025crag} integrated collaborative retrieval into conversational recommendation to inject behavioral signals for candidate generation, while Shi et al.~\cite{shi2025ocg} utilized dynamic attribute templates, open-domain retrieval, and attribute completion to construct structured profiles for fine-grained candidate comparison. Despite enriching the grounding evidence for candidates, these approaches still do not explicitly leverage preference dependencies to coordinate retrieval and candidate evaluation.

\begin{figure*}[t]
\centering
\includegraphics[width=1\textwidth]{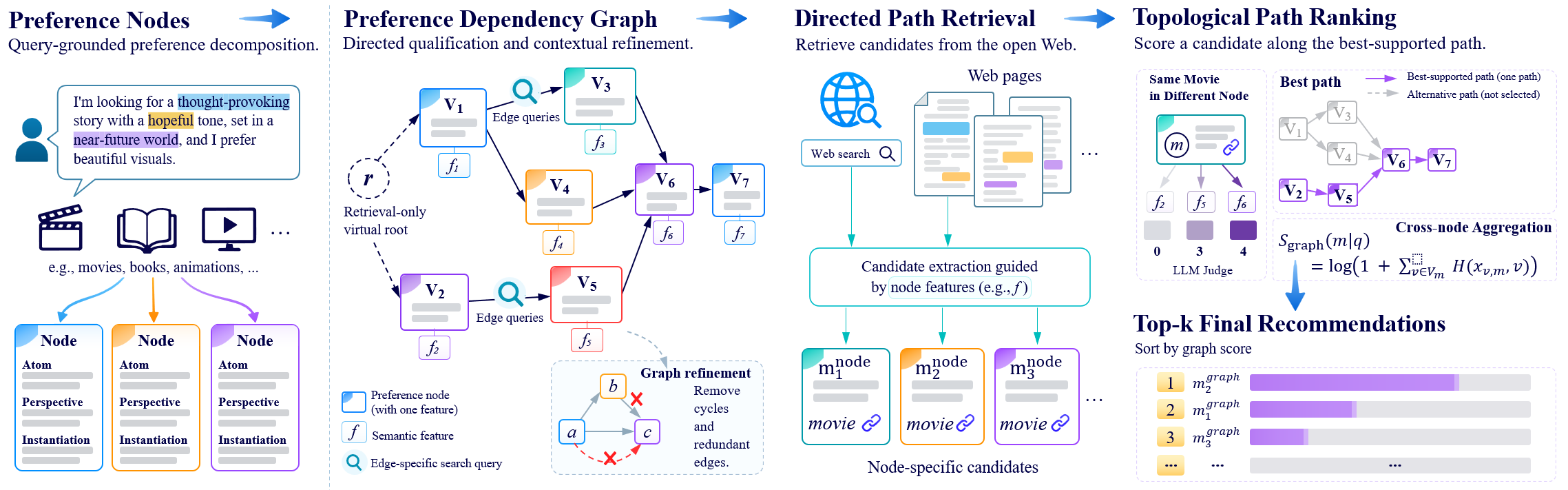}
\caption{Overview of PDG-F. PDG-A constructs the preference dependency graph, which guides directed path retrieval and topological path ranking.}
\Description{Overview of PDG-F. PDG-A constructs the preference dependency graph, which guides directed path retrieval and topological path ranking.}
\label{fig:framework}
\end{figure*}

\section{Methodology}

\subsection{Problem Formulation}
Given a user's natural language narrative request $q$, narrative-driven recommendation aims to return a ranked list of top-$K$ items from the item space $\mathcal{I}$ that satisfy the articulated needs. The request $q$ can incorporate descriptions of the target item, reference items, and user evaluations. The system output is an ordered recommendation list:
\begin{equation}
\mathcal{R}_K(q) = [i_1, i_2, \ldots, i_K], \qquad i_j \in \mathcal{I}.
\end{equation}
Notably, NDR does not presuppose a fixed candidate item pool; candidates are typically derived from web retrieval or the parametric knowledge of LLMs. Given the request $q$, the system first acquires a candidate set $\mathcal{C}_q \subseteq \mathcal{I}$ along with its supporting evidence $\mathcal{E}_q$, and subsequently scores and ranks the candidates:
\begin{equation}
(\mathcal{C}_q, \mathcal{E}_q) = \operatorname{Acquire}(q), \qquad \mathcal{R}_K(q) = \operatorname{TopK}_{i \in \mathcal{C}_q} S(i \mid q, \mathcal{E}_q),
\end{equation}
where $\operatorname{Acquire}(\cdot)$ denotes the process of obtaining candidate items and associated evidence from the aforementioned sources, and $S(i \mid q, \mathcal{E}_q)$ evaluates the extent to which candidate item $i$ satisfies request $q$ based on evidence $\mathcal{E}_q$.

\subsection{Preference Graph Construction}
PDG-A, an autonomous decision-making agent, first determines interpretive perspectives and their concrete instantiations based on narrative $q$, expanding the user's preference descriptions into a discrete node set $V_q$, where each node $v$ has an initial representation $c_v^0$. It then identifies qualification relationships between nodes within the narrative context to construct directed edges $(u,v)\in E_q$. Finally, it resolves cycles and generates a concrete preference description $f_v$ for each node in topological order, conditioned on the preferences expressed by ancestor nodes, completing the node representation as $c_v=(c_v^0,f_v)$. This yields a request-specific preference dependency graph $\mathcal{G}_q=(V_q,E_q,\mathcal{F}_q)$, where $\mathcal{F}_q=\{f_v\}_{v\in V_q}$.

\subsubsection{Node: Preference Instantiation.}
A user narrative $q$ typically intertwines referenced items, subjective feedback, and explicit criteria for recommendations. PDG-A first distills these into a set of preference atoms $\mathcal{A}_q = \{a_i\}_{i=1}^{M}$, and subsequently decomposes each atom along a two-tiered hierarchy comprising interpretive perspectives and their concrete instantiations. Formally, for each atom $a_i \in \mathcal{A}_q$:
\begin{itemize}
    \item An \textbf{interpretive perspective} $p \in \mathcal{P}_i$ specifies the semantic dimension along which the atom is interpreted (e.g., genre, director, or cast);
    \item A \textbf{concrete instantiation} $z \in \mathcal{J}_{i,p}$ anchors the specific realization within that perspective.
\end{itemize}
Each admissible tuple $(a_i, p, z)$ instantiates a distinct preference node $v \in V_q$. The initial state of node $v$ is formalized as a quadruplet:
\begin{equation}
c_v^0 = \langle \text{atom}: a_i,\, \text{perspective}: p,\, \text{instantiation}: z,\, \text{feature}: \varnothing \rangle.
\end{equation}
Here, the $\text{feature}$ field is initialized to $\varnothing$, reserved for the context-aware preference description generated later via topological propagation over ancestral constraints. 

For example, given the raw user input: ``I am looking for a movie similar to \textit{Harry Potter}\dots'', PDG-A extracts the preference atom $a_i = \text{\textit{Harry Potter}}$. Conditioned on this atom, the interpretive perspectives determined by the agent can include genre, director, or actor, among others. Taking the perspective $p = \text{actor}$ as an example, the concrete instantiation can be $z = \text{Emma Watson}$, which forms distinct nodes alongside other instances such as Daniel Radcliffe and Rupert Grint. Both the perspective set $\mathcal{P}_i$ and instantiation set $\mathcal{J}_{i,p}$ are dynamically determined by PDG-A conditioned on the current atom and the full narrative context, ultimately yielding $|V_q| = \sum_{a_i \in \mathcal{A}_q} \sum_{p \in \mathcal{P}_i} |\mathcal{J}_{i,p}|$ nodes.

\subsubsection{Edge: Qualification Dependencies.}
Independent node representations do not yet determine their semantics in a compositional context, where the same element may qualify different preference targets. To explicitly capture such relationships, we define a directed edge $(u,v)$ indicating that node $v$ imposes a concrete qualification upon the base preference expressed by node $u$. The edge direction is determined by the semantic structure of narrative $q$, reflecting a dependency directed from a foundational interpretation toward further refinement.

For any two distinct nodes, PDG-A jointly considers the complete narrative $q$, the initial representations $c_u^0, c_v^0$, and their original atoms to determine whether one node qualifies the preference target, presentation manner, or applicability conditions of the other:
\begin{equation}
d_q(u,v) \in \{(u,v),\, (v,u),\, \varnothing\},
\end{equation}
where $(u,v)$ indicates that $v$ further qualifies $u$, $(v,u)$ indicates qualification in the reverse direction, and $\varnothing$ denotes the absence of any qualification dependency. The full narrative provides the overarching context, while the two nodes specify the concrete preference interpretations under examination.

For example, if $q$ involves ``romance gradually develops through conflict'', let $u$ denote the romance node and $v$ the conflict interaction node; then $d_q(u,v) = (u,v)$, establishing the dependency $u \rightarrow v$ to signify that conflict interactions further qualify how the romance unfolds. Conversely, if $q$ merely demands romance and conflict as independent requirements, then $d_q(u,v) = \varnothing$, and no edge is created.

PDG-A evaluates all unordered node pairs and aggregates the non-empty decisions into the initial edge set:
\begin{equation}
E_q^0 = \left\{ d_q(u,v) \;\middle|\; \{u,v\} \subseteq V_q,\; u \neq v,\; d_q(u,v) \neq \varnothing \right\}.
\end{equation}
This yields the initial directed graph $\mathcal{G}_q^0 = (V_q, E_q^0)$. A node can possess multiple successors, representing divergent refinement branches of the same preference, as well as multiple predecessors, preserving its dependencies across multiple upstream contexts.

\subsubsection{Graph: Cycle Resolution.}
Local judgments between node pairs may produce directed cycles, creating circular qualifications that prevent preferences from being progressively refined in a consistent order. PDG-A therefore revisits cyclic dependencies within the complete narrative context to resolve conflicts among local judgments.

Given the current edge set $E_q^t$, we identify a strongly connected component $S_t$ containing a directed cycle. Based on narrative $q$, the relevant nodes and their original atoms, the internal dependencies, and the external connections of the component, PDG-A selects the internal edge $e_t$ least supported by the narrative for removal:
\begin{equation}
e_t\in E_q^t\cap(S_t\times S_t),
\qquad
E_q^{t+1}=E_q^t\setminus\{e_t\}.
\end{equation}
Strongly connected components are recomputed after each removal until no directed cycle remains.

For example, suppose $q$ expresses that ``romance gradually develops through conflict and deepens through subsequent reconciliation.'' Let $u=\text{romance}$, $v=\text{conflict interactions}$, and $w=\text{reconciliation after conflict}$. If local judgments establish dependencies $d_q(u,v)$, $d_q(v,w)$, and $d_q(w,u)$, they form a directed cycle $u\rightarrow v\rightarrow w\rightarrow u$. Within the narrative context, the first two dependencies progressively qualify how the romance develops, whereas $(w,u)$ returns from reconciliation to the broader romantic preference. PDG-A therefore removes $e_t=(w,u)$, retaining the refinement path $u\rightarrow v\rightarrow w$.

Once the graph is acyclic, we apply transitive reduction. If an indirect path $u\rightarrow \dots \rightarrow v$ already exists, the direct shortcut edge $(u,v)$ is removed, eliminating transitive redundancy while preserving reachability and direct refinement dependencies. This yields the final edge set $E_q$ and a valid topological ordering $\pi_q$ satisfying $\pi_q(u)<\pi_q(v)$ for all $(u,v)\in E_q$, providing a dependency-consistent order for subsequent preference propagation.

\subsubsection{Preference: Contextual Refinement.}
The initial node representation $c_v^0$ specifies a concrete direction for interpreting a preference, while the dependency structure determines its meaning within the compositional context. PDG-A completes node representations in topological order $\pi_q$, using upstream preferences as context and incorporating the current node's qualification to generate a concrete preference description.

For node $v$, let $\operatorname{Anc}(v)$ denote its ancestor set. PDG-A jointly considers the complete narrative $q$, the current representation $c_v^0$, the preference descriptions already generated for ancestor nodes in $\operatorname{Anc}(v)$, and the dependencies among these nodes to generate $f_v$. This description fills the reserved field, yielding $c_v=c_v^0[\text{feature}\leftarrow f_v]$. The topological order ensures that all ancestor descriptions are determined prior to generating $f_v$. For nodes without ancestors, the description is determined by their own initial representations and the original narrative.

For example, node $u$ represents ``a concealed motive to evade accountability behind an appearance of innocence,'' while node $v$ represents ``a lawyer's loyalty to a client and the associated moral dilemma.'' Along $u\rightarrow v$, PDG-A refines the preference at $v$ into $f_v =$ ``the lawyer's duty to protect a client may instead enable the client to evade accountability.'' The upstream concealed motive thus reshapes the interpretation of professional loyalty, explicitly manifesting the moral conflict between professional responsibility and its actual consequences.

When multiple dependency paths converge at a node, PDG-A synthesizes the path structure and the original narrative to capture the corresponding compositional or alternative relations within a single $f_v$. The resulting preference descriptions across all nodes yield the complete preference dependency graph:
\begin{equation}
\mathcal{G}_q=(V_q,E_q,\mathcal{F}_q),
\qquad
\mathcal{F}_q=\{f_v\}_{v\in V_q},
\end{equation}
providing concrete preference criteria for subsequent retrieval and candidate evaluation.

\subsection{Directed Path Retrieval}
\subsubsection{Rewriter.}
Preference dependencies represent user needs as structural relationships between nodes, whereas Web retrieval requires concrete search expressions capable of capturing these compositional qualifications. Accordingly, for each retained dependency edge $(u,v)\in E_q$, PDG-A jointly examines the complete narrative $q$ and the initial endpoint representations $c_u^0, c_v^0$ to interpret the base preference under the qualification imposed by the target node, reformulating it into an effective search query $r_{uv}$ following Google Search query syntax. Each query aligns with the language of the user narrative and translates the qualification relationship into targeted keywords and search operators, steering retrieval toward the composed preference semantics.

For a node $v$ without predecessors, its search intent is governed independently by the narrative $q$ and its own initial representation $c_v^0$. To unify query formulation, we introduce a virtual root node $\mathrm{ROOT}$ and generate an entry query $r_{\mathrm{ROOT},v}$ for each such predecessor-free node. Let $E_q^+$ denote the edge set augmented with these virtual entry edges. The resulting queries collectively constitute the retrieval query set:
\begin{equation}
\mathcal{Q}_q=\{r_{uv}\mid(u,v)\in E_q^+\}.
\end{equation}

\subsubsection{Query.}
We execute the query set $\mathcal{Q}_q$ through the Serper API and collect URLs from the returned organic search results. We filter these URLs by resource type, removing invalid URLs, social media websites, PDF documents, videos, and other excluded resource types.

Let $U(r_{uv})$ denote the filtered URL set returned for query $r_{uv}$. Through the dependency edge $(u,v)\in E_q^+$, these URLs are associated with the target node $v$, providing sources for subsequent candidate recall under its specific preference. Each node aggregates and deduplicates the URLs returned by all its incoming-edge queries:
\begin{equation}
\mathcal{U}_v = \bigcup_{u:\, (u,v)\in E_q^+} U(r_{uv}).
\end{equation}
The same URL may be associated with multiple nodes, allowing candidate items and relevant descriptions to be extracted separately according to each node's preference.

\subsubsection{Recall.}
For the URL set $\mathcal{U}_v$ associated with node $v$, we fetch the webpage text, resolve redirection targets, and inspect MIME resource types, pruning failed requests, empty bodies, and exceptionally large pages exceeding 200,000 characters. For each node, we then select at most the top 3 pages sorted by text length in descending order, and segment the contents into chunks of up to 12,000 characters with a 2,000-character overlap between adjacent passages, forming the text corpus $\mathcal{B}_v$. When identical URLs are shared across multiple nodes, the parsed text and chunks are reused while preserving their individual node assignments.

For each chunk $b \in \mathcal{B}_v$, the model extracts candidate items $i$ along with their associated textual evidence $x_{v,i}^{(b)}$, guided by the node's refined preference description $f_v$. Each extracted entry is formalized as $\langle \text{name}, \text{year}, x_{v,i}^{(b)} \rangle$, recording the item name, release year if explicitly mentioned (or empty otherwise), and concrete descriptions pertinent to the active preference. A candidate record is formed only when a chunk simultaneously provides both unambiguous item identification and substantive content. Crucially, the extraction faithfully retains statements whether they support or contradict the preference, decoupling evidence extraction from subsequent preference verification. Because extraction is strictly conditioned on node $v$, extracted candidates and their evidence naturally inherit node provenance, allowing a single text chunk to yield distinct records under different node specifications $f_v$.

Following extraction, we identify and exclude reference items already mentioned in the initial request $q$ using name and year metadata, and then deduplicate candidates within each node. For the same item, identical descriptions are removed, and distinct descriptions are sequentially concatenated into a consolidated text $x_{v,i}$, with their source URLs retained, yielding the node candidate set $\mathcal{C}_v$. Cross-node descriptions are maintained independently to ensure that the factual evidence for an item across different preference facets remains unentangled. This produces the global candidate set and the node-grounded evidence set for the narrative request:
\begin{equation}
\mathcal{C}_q = \bigcup_{v \in V_q} \mathcal{C}_v,
\qquad
\mathcal{E}_q = \left\{ (v, i, x_{v,i}) \;\middle|\; v \in V_q,\; i \in \mathcal{C}_v \right\},
\end{equation}
where $x_{v,i}$ encapsulates the web-grounded descriptions aligned with preference $f_v$, providing fine-grained evidence for candidate evaluation along the preference dependency paths.

\subsection{Topological Path Ranking}
\subsubsection{Initial Scoring.}
We collect the preference descriptions of the current node $v$ and all its ancestors into $\mathcal{F}_v=\{f_u\mid u\in\operatorname{Anc}(v)\cup\{v\}\}$, where $\operatorname{Anc}(v)$ denotes the ancestor set of $v$, and \(f_u\) is the concrete preference description of node \(u\). For each item $i$ recalled at node $v$, we hold its retrieved description $x_{v,i}$ fixed and compare it against every preference description $f_u\in\mathcal{F}_v$, assessing how the same textual evidence supports the active and upstream preferences.

PDG-A assigns a relation label based solely on the text pair $(x_{v,i},f_u)$, which the system then maps to an initial score:
\begin{equation}
s(x_{v,i},f_u)=
\begin{cases}
0, & \text{explicit contradiction},\\
1, & \text{weak but meaningful relation},\\
2, & \text{relation through reasonable inference},\\
3, & \text{explicit support},\\
4, & \text{paraphrase}.
\end{cases}
\label{eq:initial-scoring}
\end{equation}
The judgment respects the qualification scope, preference strength, and permitted alternatives expressed in $f_u$. Missing details alone do not establish an explicit contradiction, and permitted alternatives need not hold simultaneously. An explicit contradiction assigns zero only to the current pairwise comparison, without removing the candidate or altering its other comparison scores. These initial scores provide the foundation for subsequent dependency-path computation.

\begin{table*}[t]
\centering
\caption{Main comparison results. Best values are bold; second-best values are underlined. For ties, only the last occurrence in each column is marked.}
\label{tab:main-results}

\begin{tabular*}{\textwidth}{@{\extracolsep{\fill}}l*{9}{c}@{}}
\toprule
Method & \multicolumn{3}{c}{Movie} & \multicolumn{3}{c}{Book} & \multicolumn{3}{c}{Anime} \\
\cmidrule(lr){2-4}\cmidrule(lr){5-7}\cmidrule(lr){8-10}
& P@10 & R@10 & NDCG@10 & P@10 & R@10 & NDCG@10 & P@10 & R@10 & NDCG@10 \\
\midrule
qwen3.8-27B & 0.2730 & 0.1645 & 0.3280 & 0.1100 & 0.0989 & 0.1492 & 0.0267 & 0.0383 & 0.0429 \\
minimax-m3 & 0.3070 & 0.1885 & 0.3647 & 0.1900 & 0.1589 & 0.2606 & 0.0633 & 0.0791 & 0.0979 \\
deepseek-v4-pro & 0.3170 & \underline{0.1954} & 0.3753 & 0.1867 & 0.1648 & 0.2526 & 0.0567 & 0.0736 & 0.0884 \\
\midrule
OCG-Agent & 0.2410 & 0.1448 & 0.2647 & 0.2200 & 0.2299 & 0.3264 & 0.1133 & 0.1356 & 0.1901 \\
OCG-Agent (+Year) & 0.2880 & 0.1693 & 0.3284 & \underline{0.2200} & \underline{0.2299} & \underline{0.3264} & 0.1167 & 0.1389 & 0.1922 \\
Perplexity AI & 0.2700 & 0.1656 & 0.3206 & 0.1733 & 0.1396 & 0.2182 & \underline{0.1667} & \underline{0.1692} & 0.2028 \\
GPT-6-Astra (Codex) & \underline{0.3210} & 0.1938 & \underline{0.3771} & 0.1800 & 0.1410 & 0.2078 & 0.1467 & 0.1411 & 0.1888 \\
\midrule
Ours & 0.3010 & 0.1865 & 0.3631 & 0.4233 & 0.3480 & 0.5101 & 0.2467 & 0.2722 & \underline{0.3025} \\
Ours (+Year) & \textbf{0.3650} & \textbf{0.2229} & \textbf{0.4511} & \textbf{0.4233} & \textbf{0.3480} & \textbf{0.5101} & \textbf{0.2467} & \textbf{0.2722} & \textbf{0.3085} \\
\bottomrule
\end{tabular*}
\end{table*}

\subsubsection{Topological Path Computation.}
The initial scores quantify how an item description supports individual node preferences. We further aggregate these scores along dependency paths to assess how the same description supports progressively refined preferences. For an item $i$ recalled at node $v$, we hold its retrieved description $x_{v,i}$ fixed and traverse the subgraph induced by node $v$ and all its ancestors $\operatorname{Anc}(v)$ according to the topological ordering $\pi_q$. Let $\operatorname{Pa}(u)=\{p\mid(p,u)\in E_q\}$ denote the immediate predecessor set of node $u$, and let $H(x_{v,i},u)$ denote the optimal path score ending at $u$, computed using this fixed description. The recurrence is defined as:
\begin{equation}
H(x_{v,i},u)=
\begin{cases}
s(x_{v,i},f_u),
& \operatorname{Pa}(u)=\varnothing,\\[2pt]
s(x_{v,i},f_u)
+\gamma\displaystyle\max_{p\in\operatorname{Pa}(u)}
H(x_{v,i},p),
& \text{otherwise},
\end{cases}
\label{eq:ranking-equation}
\end{equation}
where $s(x_{v,i},f_u)$ is the initial score defined in Eq.~\eqref{eq:initial-scoring}, and $\gamma\in(0,1]$ is a discount factor (set to $0.7$ in our implementation). A node without predecessors initializes its path score with its own initial score. For a node with predecessors, its initial score is combined with the discounted optimal path score from its predecessors. The topological order guarantees that all predecessor path scores are determined before evaluating node $u$.

When multiple predecessors converge at a node, the maximum operator isolates the highest-scoring refinement path. Along this trajectory, each additional edge traced upstream introduces an extra factor of $\gamma$, exponentially decaying the contribution of more distant ancestors while prioritizing preferences proximal to the current node. The recurrence consistently evaluates the identical evidence $x_{v,i}$ and ultimately yields $H(x_{v,i},v)$, the path score for item $i$ under node $v$. The virtual root used for retrieval is excluded from scoring.

The same candidate item may be recalled across multiple nodes with distinct node-specific descriptions. Let $\mathcal{V}_i=\{v\in V_q\mid i\in\mathcal{C}_v\}$ denote the set of nodes where item $i$ is retained. We aggregate its path scores across these nodes and apply a logarithmic compression to obtain the final graph-grounded score:
\begin{equation}
S_{\mathrm{graph}}(i\mid q)
=
\log\left(
1+\sum_{v\in\mathcal{V}_i}H(x_{v,i},v)
\right).
\end{equation}
Each node evaluates its path score independently using its own extracted description, and evidence from multiple webpages mapped to the same node is consolidated to contribute only once. An ancestor preference participating in the path computation of a descendant candidate does not spuriously introduce an extra candidate record at that ancestor node.

\section{Experiment}
\subsection{Experimental Setup}
\paragraph{Datasets.}
Following the experimental setup of OCG-Agent~\cite{shi2025ocg}, we first adopt Reddit MovieSuggestions, a widely used benchmark for narrative-driven recommendation~\cite{eberhard2019evaluating,eberhard2020embedding,eberhard2024computing,eberhard2025llmndr,shi2025ocg}. We seek to further validate the capability of PDG-F across diverse tasks. Given that the proprietary \emph{education}~\cite{shi2025ocg} dataset is not publicly accessible, we instead collect two datasets from public online communities, namely \emph{Book} with 30 English queries from Reddit and \emph{Anime} with 30 Chinese queries from Xiaohongshu and Zhihu. 

We adapt the curation methodology of Reddit MovieSuggestions~\cite{eberhard2019evaluating} to construct these two benchmarks. For each sample, the input query is obtained by faithfully summarizing an informative recommendation post while preserving user preferences, constraints, and contextual details. The ground-truth items are scored based on community support, defined by the peak engagement achieved across recommending replies, namely the net vote score on Reddit or the like count on Xiaohongshu and Zhihu. When reliable engagement scores are unavailable for a query, we quantify community support by recommendation frequency within the collected replies.

\paragraph{Evaluation.}
We evaluate recommendation performance using Precision@10, Recall@10, and NDCG@10~\cite{jarvelin2002cumulated,manning2008ir}. For each query, we follow the original setting of OCG-Agent~\cite{shi2025ocg} by performing retrieval once and running RankGPT reranking~\cite{sun2023rankgpt} three times, reporting the averaged results. For our method, PDG-F, ranking scores are computed deterministically from the constructed graph and collected evidence, so a single ranking run is sufficient.

\paragraph{Baselines.}
Given the continuing improvements in LLM capabilities, we rerun experiments with frontier models, including three leading Chinese models, Qwen-3.8-27B, DeepSeek-V4-Pro, and MiniMax-M3, as well as GPT-6-Astra, a leading international model. We also include Perplexity, a widely used commercial AI search engine. We set the reasoning effort to \texttt{high} for all models except GPT-6-Astra, which uses \texttt{xhigh}. For OCG-Agent~\cite{shi2025ocg}, we reproduce its pipeline using
Qwen-3.8-27B as the LLM backbone. Our method, PDG-F, uses the same Qwen-3.8-27B backbone, which we deploy locally on four H20 GPUs with 140\,GB of memory each. Since the remaining model API calls and search services incur monetary costs, we follow OCG-Agent~\cite{shi2025ocg} and randomly sample 100 Movie queries for evaluation.

\begin{figure*}[t]
\centering
\includegraphics[width=0.99\textwidth]{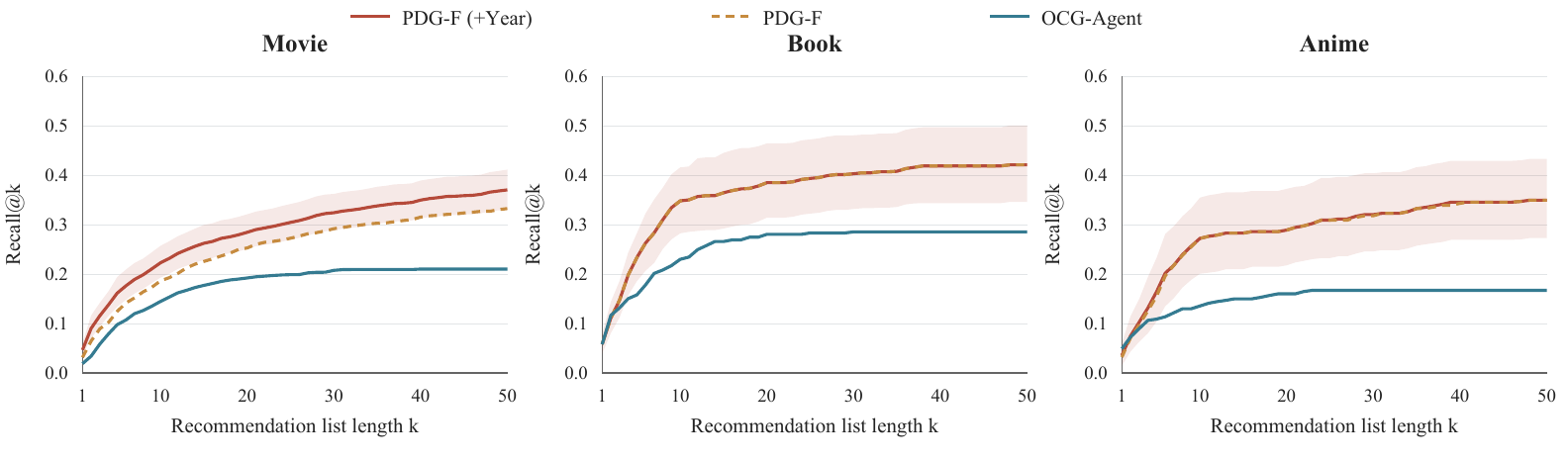}
\caption{Recall at cutoffs $k \in [1, 50]$. Shaded bands denote 95\% bootstrap confidence intervals. The dashed PDG-F curves largely overlap with the solid PDG-F (+Year) curves on Book and Anime.}
\Description{Recall at cutoffs $k \in [1, 50]$. Shaded bands denote 95\% bootstrap confidence intervals. The dashed PDG-F curves largely overlap with the solid PDG-F (+Year) curves on Book and Anime.}
\label{fig:recall-topk}
\end{figure*}

\subsection{Result Analysis}
\paragraph{Overall Performance.}
Table~\ref{tab:main-results} summarizes the performance comparison across the three benchmark datasets. Overall, our proposed PDG-F (+Year) achieves the best performance on all three datasets across all evaluation metrics. Specifically, on the Movie dataset, PDG-F (+Year) outperforms the strongest competing baselines by 13.7\%, 14.1\%, and 19.6\% in P@10, R@10, and NDCG@10, respectively. The gains are even larger on the Book and Anime benchmarks, where PDG-F (+Year) achieves a 92.4\% improvement in P@10 on Book and a 60.9\% gain in R@10 on Anime over the best competing methods. These consistent improvements across diverse domains substantiate the value of explicitly modeling preference dependencies and integrating supporting evidence for narrative-driven recommendation.

\paragraph{Further Analysis.}
The performance disparities across domains and model variants may partly reflect temporal retrieval biases and differences in knowledge coverage. First, web search often favors recent and popular content, whereas the Movie benchmark frequently targets historical or classic titles. 
Accordingly, the (+Year) variant applies rule-based temporal comparison and filtering using the extracted release year metadata to exclude candidates that violate temporal constraints.
Incorporating year information helps counter this mismatch by reducing the prominence of newer candidates, yielding substantial gains on Movie while gains on Anime remain more limited. Second, classic movie plots and item profiles are likely well represented in LLM pretraining corpora, helping direct prompting remain competitive on Movie. Newer or less documented content in Book and Anime may be less well covered by parametric knowledge, making external retrieval more valuable in these domains.

\section{Ablation Study}

\subsection{Candidate Recall Analysis}
We compare the candidate acquisition of PDG-F and OCG-Agent across overall pool coverage, relevant item overlap, and ranking-level Recall@$k$.

\subsubsection{Candidate Coverage.}
Table~\ref{tab:candidate-recall} compares candidate acquisition against the strongest retrieval baseline, OCG-Agent. Across the three datasets, PDG-F constructs substantially broader candidate pools, averaging roughly 250 to 690 items per request, compared with around 20 for OCG-Agent. Crucially, this expansion does not merely introduce noise, as pool recall approximately doubles or more across the three domains, increasing from 0.2102 to 0.4858 on Movie, from 0.2854 to 0.5481 on Book, and nearly tripling from 0.1672 to 0.4734 on Anime. This confirms that decomposing narrative preferences across dependency nodes effectively retrieves diverse, high-utility ground-truth items, establishing a much stronger candidate foundation for subsequent topological ranking. Importantly, because PDG-F scores candidates through deterministic graph accumulation rather than prompting-based listwise rerankers such as RankGPT, such candidate pool expansion avoids the pitfalls of context window limits~\cite{qin2024pairwise}, position bias~\cite{tang2024middle}, and prohibitive inference latency.

\begin{table}[h]
\centering
\caption{Candidate-pool size and relevant-item coverage.}
\label{tab:candidate-recall}
\begin{tabular*}{\columnwidth}{@{\extracolsep{\fill}}llrr@{}}
\toprule
Dataset & Method & \shortstack[r]{Avg.\\candidates} & \shortstack[r]{Pool\\recall} \\
\midrule
Movie & PDG-F (+Year) & 251.84 & 0.4858 \\
& OCG-Agent & 22.52 & 0.2102 \\
\midrule
Book & PDG-F (+Year) & 687.33 & 0.5481 \\
& OCG-Agent & 17.63 & 0.2854 \\
\midrule
Anime & PDG-F (+Year) & 366.30 & 0.4734 \\
& OCG-Agent & 17.33 & 0.1672 \\
\bottomrule
\end{tabular*}
\end{table}

\subsubsection{Recall across Recommendation Lengths.}
Figure~\ref{fig:recall-topk} illustrates how broader candidate acquisition translates into ranking recall across varying recommendation lengths. As $k$ grows from 1 to 50, the recall trajectories exhibit markedly different saturation behaviors. Due to its constrained candidate pool, OCG-Agent rapidly plateaus around $k=20$ across all datasets, yielding minimal gains as the list extends further. In contrast, PDG-F with year maintains a continuous upward trajectory, consistently surfacing additional relevant items throughout longer lists. Remarkably, its Recall@10 alone surpasses the entire complete-pool recall of OCG-Agent across all three benchmarks. On Anime, OCG-Agent exhibits a slight advantage at the extreme top position, with a Recall@1 of 0.0492 versus 0.0359 for PDG-F with year, yet PDG-F rapidly overtakes it as $k$ reaches 10 and beyond. This demonstrates that the candidate coverage advantages of PDG-F become increasingly pronounced when presenting multi-item recommendation lists.

\subsubsection{Complementarity of Retrieved Items.}
\begin{figure}[h]
\centering
\includegraphics[width=0.99\columnwidth]{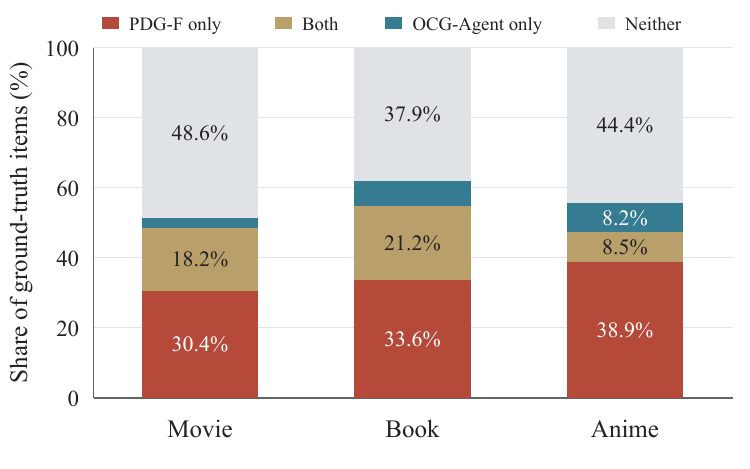}
\caption{Ground truth item overlap between candidate pools.}
\Description{Ground truth item overlap between candidate pools.}
\label{fig:recall-evidence}
\end{figure}

Figure~\ref{fig:recall-evidence} further examines where the additional coverage comes from. Relevant items found exclusively by PDG-F account for 30.41\%, 33.57\%, and 38.85\% of all ground-truth items on Movie, Book, and Anime, respectively, while those found exclusively by OCG-Agent account for 2.86\%, 7.29\%, and 8.24\%. These proportions are computed per query and then macro-averaged. Thus, PDG-F's coverage advantage includes a substantial number of relevant works absent from the baseline pools, illustrating the additional coverage obtained from candidate acquisition under different preference perspectives. OCG-Agent also retrieves some relevant items missed by PDG-F, indicating that the candidate sources remain partially complementary.

\subsection{Ranking Analysis}
We analyze node aggregation strategies, score discriminability, oracle performance gaps, and the role of dependency modeling.

\begin{table}[h]
\centering
\caption{Evaluation of candidate ranking strategies.}
\label{tab:ranking-strategies}
\begin{tabular*}{\columnwidth}{@{\extracolsep{\fill}}lrrr@{}}
\toprule
Strategy & Movie & Book & Anime \\
\midrule
Random & 0.0641 & 0.0163 & 0.0242 \\
Mean node & 0.1506 & 0.0178 & 0.0667 \\
Max node & 0.3205 & 0.1997 & 0.1255 \\
\midrule
Full strategy & \textbf{0.4511} & \textbf{0.5101} & \textbf{0.3085} \\
\bottomrule
\end{tabular*}
\end{table}

\subsubsection{Ranking Strategy Comparison.}
Table~\ref{tab:ranking-strategies} compares node aggregation strategies on identical candidate pools against a random ranking baseline. Mean aggregation degrades severely across all datasets because averaging dilutes decisive signals across nodes. Taking the maximum node score provides a viable baseline, confirming that prominent local matches offer useful discrimination. Notably, this maximum node score already benefits from our propagation strategy by inheriting contextual preference constraints from ancestor nodes. Even with this contextual advantage, the full additive strategy consistently achieves higher NDCG@10 across all domains, and even more notably, it more than doubles the performance on Book and Anime, demonstrating the effectiveness of our cumulative evidence design.

\begin{figure}[h]
\centering
\includegraphics[width=0.99\columnwidth]{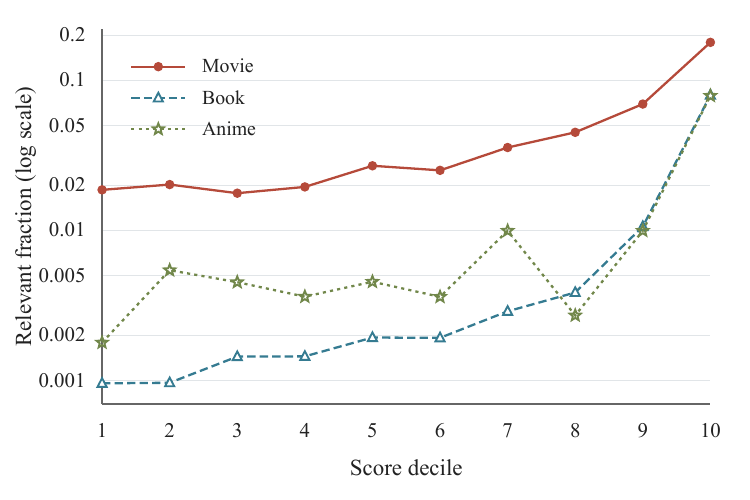}
\caption{Relevance rates across score deciles.}
\Description{Relevance rates across score deciles.}
\label{fig:score-relevance}
\end{figure}

\subsubsection{Graph Scores and Relevance.}
Figure~\ref{fig:score-relevance} evaluates the discriminative power of the computed graph scores within the expanded candidate pools. Partitioning candidates into score deciles reveals that ground-truth relevant items are heavily concentrated in the top decile, achieving relevance rates of 17.97\%, 8.01\%, and 7.92\% on Movie, Book, and Anime, respectively, compared with only 1.87\%, 0.10\%, and 0.18\% in the bottom decile. This sharp separation directly reflects the scoring quality because candidates assigned high graph scores are significantly more likely to be truly relevant, whereas irrelevant candidates are successfully suppressed into lower deciles. The cross-domain consistency of this trend confirms that the graph score functions as a reliable relevance indicator, validating that evaluating evidence along interdependent preference paths effectively captures whether an item satisfies the user's composite needs.

\begin{figure}[h]
\centering
\includegraphics[width=0.99\columnwidth]{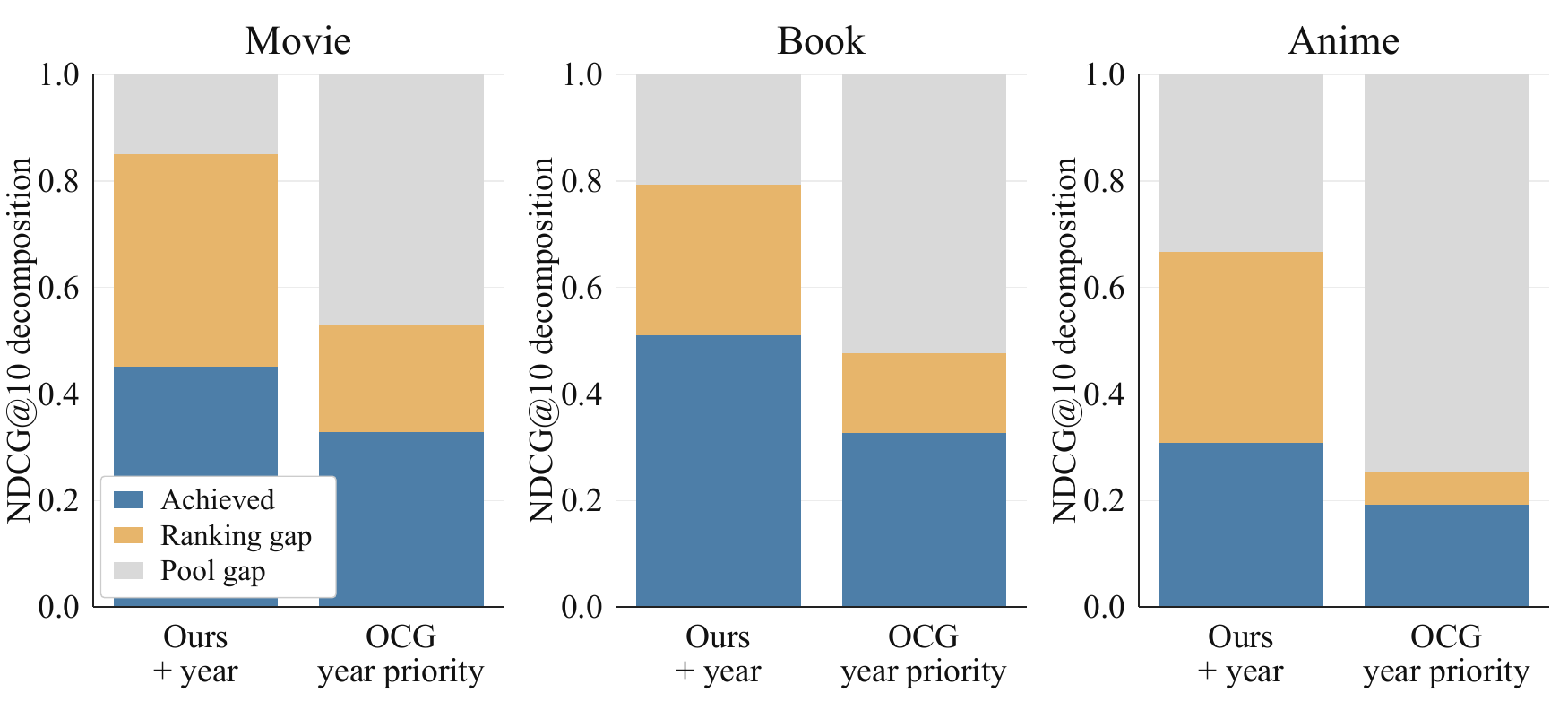}
\caption{Decomposition of ideal NDCG@10 into pool and ranking gaps.}
\Description{Decomposition of ideal NDCG@10 into pool and ranking gaps.}
\label{fig:oracle-decomposition}
\end{figure}

\subsubsection{Candidate Acquisition and Ranking Gaps.}
We finally examine Figure~\ref{fig:oracle-decomposition} to decompose the gap to ideal NDCG@10 into candidate-pool coverage and within-pool ranking accuracy. Across all domains, PDG-F achieves markedly higher pool-oracle upper bounds than OCG-Agent, confirming that multi-path acquisition builds a far stronger candidate foundation. At the same time, the visible ranking gaps highlight that while candidate retrieval raises the performance ceiling, ranking retrieved relevant items into top positions remains the primary bottleneck and leaves substantial room for future improvement.

\subsubsection{Dependency Modeling Analysis.}
The empirical comparison between PDG-F and OCG-Agent naturally constitutes an ablation over dependency modeling. Specifically, our preference nodes refine the dynamic attribute representations of OCG-Agent through redesigned extraction sources and field specifications, while the overarching innovation lies in establishing directed qualification graphs. In the absence of directional dependency edges, candidate acquisition reduces to uncoordinated retrieval across isolated preference nodes. Furthermore, lacking the underlying graph structure precludes our topological path ranking, forcing candidate evaluation to revert to standard prompting-based reranking via RankGPT. Under this double degradation, the framework essentially degenerates into the baseline pipeline of OCG-Agent. The substantial performance margins across all evaluation domains, as shown in Table~\ref{tab:main-results}, therefore confirm that directional qualification modeling serves as the decisive driver for enhanced candidate acquisition and relevance discrimination.

\section{Conclusion}
We presented PDG-F, a framework for narrative-driven recommendation that captures how user preferences qualify one another through directed acyclic graphs. By carrying these qualification relationships into candidate acquisition and topological evidence accumulation, the framework connects narrative understanding with evidence-based recommendation. Experiments across three domains demonstrate broader relevant-item coverage and improved recommendation performance over competitive baselines. Oracle analysis further shows that this broader coverage raises the attainable performance ceiling, while the remaining ranking gap motivates more accurate evidence scoring to realize that potential.

\bibliographystyle{ACM-Reference-Format}
\bibliography{references}

\end{document}